# Nanometer-scale operando electric field mapping in oxide junctions by correlative STEM-EBIC and 4D-STEM

Yining Xie[†], Eoin Moynihan[†], Xingyao Li[‡§], Richard Beanland[†], Aditya Singh[†], Marin Alexe[†], Mingmin Yang[‡§], Ana Sanchez[*†]

*Corresponding Author Email: A.M.Sanchez@warwick.ac.uk

† Department of Physics, University of Warwick, Coventry, CV4 7AL, United Kingdom
‡ School of Emerging Technology, The University of Science and Technology of China, Hefei, 230026, China
§ Hefei National Laboratory, Hefei, Anhui, 230088, China

**Abstract**

Understanding local electric fields at oxide interfaces is essential for linking interfacial electrostatics to device functionality, yet conventional electrical measurements infer these fields only indirectly. Here, we combine operando scanning transmission electron microscope electron-beam-induced current (STEM-EBIC) with four-dimensional-STEM (4D-STEM) to quantitatively reconstruct the bias-dependent electric field spatial distribution across an oxide Schottky junction at nanometer scale. Using $La_{0.67}Sr_{0.33}MnO_3$/Nb:$SrTiO_3$ (LSMO/NSTO) as a model system, we fabricate an electron-transparent junction on a MEMS biasing platform and verify that it retains the rectifying transport behavior. STEM-EBIC provides complementary information addressing several key limitations in quantitative 4D-STEM field mapping for heterojunctions. The reconstructed electric field profiles show a pronounced nonlinear decay within the depletion region and an extended penetration into the LSMO. These results directly reveal the deviations of oxides Schottky junction from the ideal conventional Schottky depletion model, providing experimental signatures of non-classical interfacial electrostatics. Our correlative approach enables quantitative nanoscale electric field mapping in operating oxide heterojunctions, providing a basis for linking interfacial electrostatics to macroscopic transport and guiding oxide devices design.

## 1. INTRODUCTION

Oxide heterostructures host emergent transport behaviors that are often strongly coupled to local structure and can be tuned and vary across interfaces or domain walls over only a few nanometers, or even a few unit cells(*1-3*). For example, $SrTiO_3$(STO)-based Schottky heterojunctions exhibit rectification behavior that can depart markedly from conventional Schottky devices, including anomalous temperature-dependent rectification and capacitance-voltage ($C$-$V$) characteristics(*4, 5*). However, conventional electrical measurements primarily probe the overall junction response and the underlying field distribution can only be inferred indirectly. Understanding these behaviors requires the local electric fields to be spatially mapped and ideally, while the device is operating. Such measurements are particularly challenging for STO-based heterojunctions because the depletion region can be confined to only a few tens of nm(*6*). While this extreme spatial localization makes oxide heterostructures especially attractive platforms for next generation novel electronic devices, it also presents a significant resolution challenge for the direct mapping of local electric fields(*3, 7*).

In this context, (scanning) transmission electron microscopy ((S)TEM) has emerged as a powerful platform for probing local structure and electrostatic behavior with sub-angstrom resolution(*8, 9*). Annular dark-field (ADF) and (annular) bright field imaging can resolve atomic structure, while energy-dispersive X-ray spectroscopy (EDX) and electron energy-loss spectroscopy (EELS) provide complementary information on local composition and electronic structure(*10-13*). Electrostatic fields can be investigated using differential phase contrast (DPC) or electron holography(*14, 15*). More recently, pixelated detectors have enabled four-dimensional STEM (4D-STEM), allowing projected electric fields to be mapped through analysis of center-of-mass (CoM) caused by the deflection of the electron beam(*16, 17*). In parallel, MEMS-based biasing holders have enabled operando electrical measurements inside TEM. This has transformed TEM into a multifunctional platform where electrical properties can be directly correlated with the structure. These developments have also opened new opportunities for electron-beam-induced current (EBIC) measurements in STEM (STEM-EBIC). STEM-EBIC enables the characterization of depletion width, carrier diffusion length, recombination dynamics, and electrically active defects(*18, 19*). Compared with conventional EBIC in scanning electron microscopy (SEM-EBIC), STEM-EBIC provides higher spatial resolution and a geometry particularly well suited for investigating buried interfaces and junctions, making it especially attractive for oxide heterostructures(*3, 20*).

Although recent studies have demonstrated the feasibility of these operando electrical measurements(*21-23*), several important challenges remain for quantitative electric field mapping. First, TEM sample preparation produces electrically inactive surface layers on both sides of the MEMS device, so the electrically active thickness ($t_{eff}$) may differ from the physical specimen thickness and must be accounted for when converting CoM deflection to electric field(*24*). Second, changes in the mean inner potential (MIP) across a heterojunction introduce additional contributions to CoM deflection, which must be separated from the electrostatic field of interest(*25, 26*). Third, the small convergence semi-angle required for 4D-STEM field mapping produces a finite probe size of several nanometers, broadening the measured field profile and complicating direct interpretation of the underlying field distribution(*25, 27*).

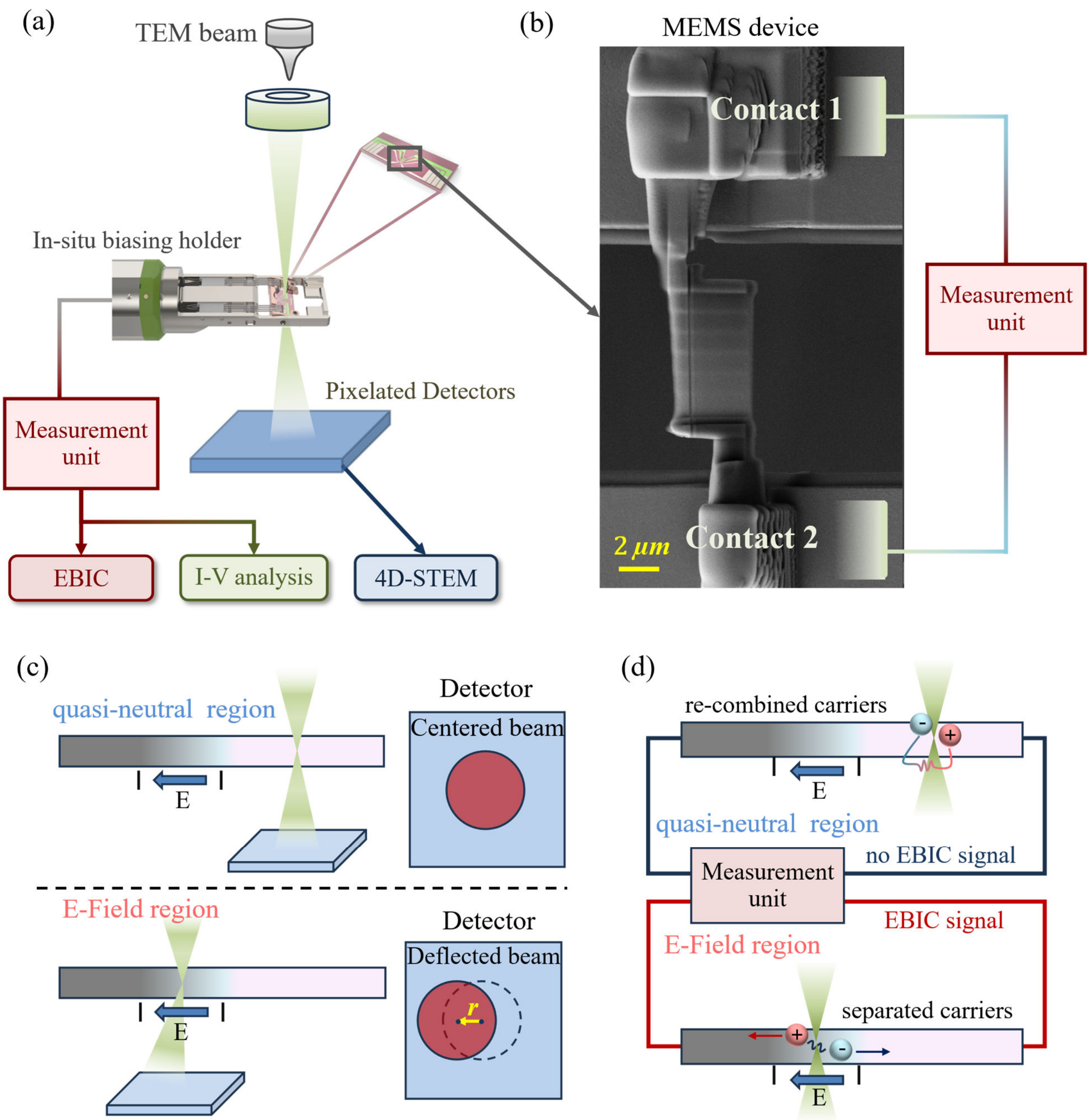

**Figure 1 In-situ experimental configuration.** (a) Schematic of the in-situ STEM setup, where pixelated detectors acquire 4D-STEM datasets, while a biasing holder applies external bias and collects EBIC signal; (b) SEM image of the device geometry; (c) schematic diagram of 4D-STEM measuring internal electric field from a quasi-neutral region without an electric field (top) and under electric field presenting (bottom); (d) schematic diagram of STEM-EBIC from a quasi-neutral region (top) and under electric field presenting (bottom).

Here, we establish an operando TEM approach that combines STEM-EBIC and 4D-STEM to quantitatively probe local electric fields in an operating oxide heterojunction under applied bias.

We use $La_{0.67}Sr_{0.33}MnO_3/Nb:SrTiO_3$ Schottky heterojunction device (LSMO/NSTO) as a model system, constituted by the spin-polarized half-metal LSMO and *n*-semiconductor NSTO with rectifying transport behavior and a highly localized junction electric field(*28, 29*). This interface has attracted considerable attention for oxide electronics and spintronics, including spin-polarized junctions and resistive switching devices(*29*). We first demonstrate that the TEM-integrated device preserves its characteristic rectifying behavior under operando conditions. We then combine STEM-EBIC and 4D-STEM as complementary techniques to measure the electric field distribution. STEM-EBIC provides independent measurements including the establishment of a near-zero-field condition to minimize the influence of MIP, and $t_{eff}$ to overcome the limitations in 4D-STEM for oxide heterojunctions.

Using this correlative approach, we reconstruct, for the first time, the bias dependent electric field distribution across an operating oxide Schottky junction device with nanometer scale resolution. The reconstructed 1D electric field line profiles at different bias shows a nonlinear decay of the electric field within the NSTO depletion region, together with an extended penetration into the LSMO. These findings directly demonstrate the deviation of electric field distribution at oxide Schottky junction devices from the conventional semiconductor picture. More importantly, our approach provides a feasible route for measuring electric field distribution at operating oxide devices at nanometer scale.

## 2. RESULTS

### 2.1 Device geometry and IV characteristics

Fig. 1a illustrates the operando STEM platform used in this work, in which a nanoscale device is integrated onto a MEMS biasing holder using a focused ion beam (FIB) microscope. The device geometry is shown in Fig. 1b. The Pt top electrode and NSTO bottom substrate are connected to the two contacts of the MEMS biasing holder, which is then connected to an electrical analysis system. This configuration enables electrical measurements, external bias and EBIC signal collection during correlative STEM measurements. As shown in fig. 1c-d, 4D-STEM measures the projected electric field through shifts in the CoM of the diffraction patterns from the electron beam deflection by the electric field, whereas STEM-EBIC detects current arising from the separation of electron-hole pairs by the electric field. Detailed device fabrication procedures, measurement principles and analysis procedures are described in the *Materials and Methods* section.

A magnified view of the device structure is given in fig. 2a. The atomic-resolution ADF-STEM image in fig. 2b confirms the atomically sharp interface of epitaxial LSMO/NSTO interface. LSMO/NSTO interface, constituting a metal/*n*-semiconductor Schottky junction. Additional characterization of the structure and composition, including EDX analysis, is presented in the *Materials and Methods* section and figs. SI1-3.

Before performing the operando STEM measurements, it is important to confirm that the MEMS device remains electrically operational after FIB preparation and integration. Here, the current density-voltage ($J$-$V$) characteristics serve as the primary metric for evaluating the electrical performance of the LSMO/NSTO Schottky junction and form the basis for the subsequent *in-situ* electric field analysis. Fig. 2c compares the $J$-$V$ characteristics of the TEM device with those of the corresponding macroscopic (bulk) device (from -1.5V to 1.5V, as described in *Materials and Methods* section). Both devices exhibit pronounced rectifying behavior, confirming the formation of a Schottky barrier at the oxide interface, and validating the successful device integration in the MEMS chip. For the macroscopic device, transport parameters were extracted combining $J$-$V$ and $C$-$V$ measurements. For the MEMS device where a $C$-$V$ measurement is not appliable, the Cheung-Cheung method is employed to extract fitting parameters for the $J$-$V$ curves(*30*). The extracted Schottky barrier height ($\varphi_B$), ideality factor ($n$), depletion width at zero bias($W_0$), and series ($R_s$) and parallel resistances ($R_p$) are summarized in Table 1. Detailed procedures are provided in *Materials and Methods* and Figs. SI4-5.

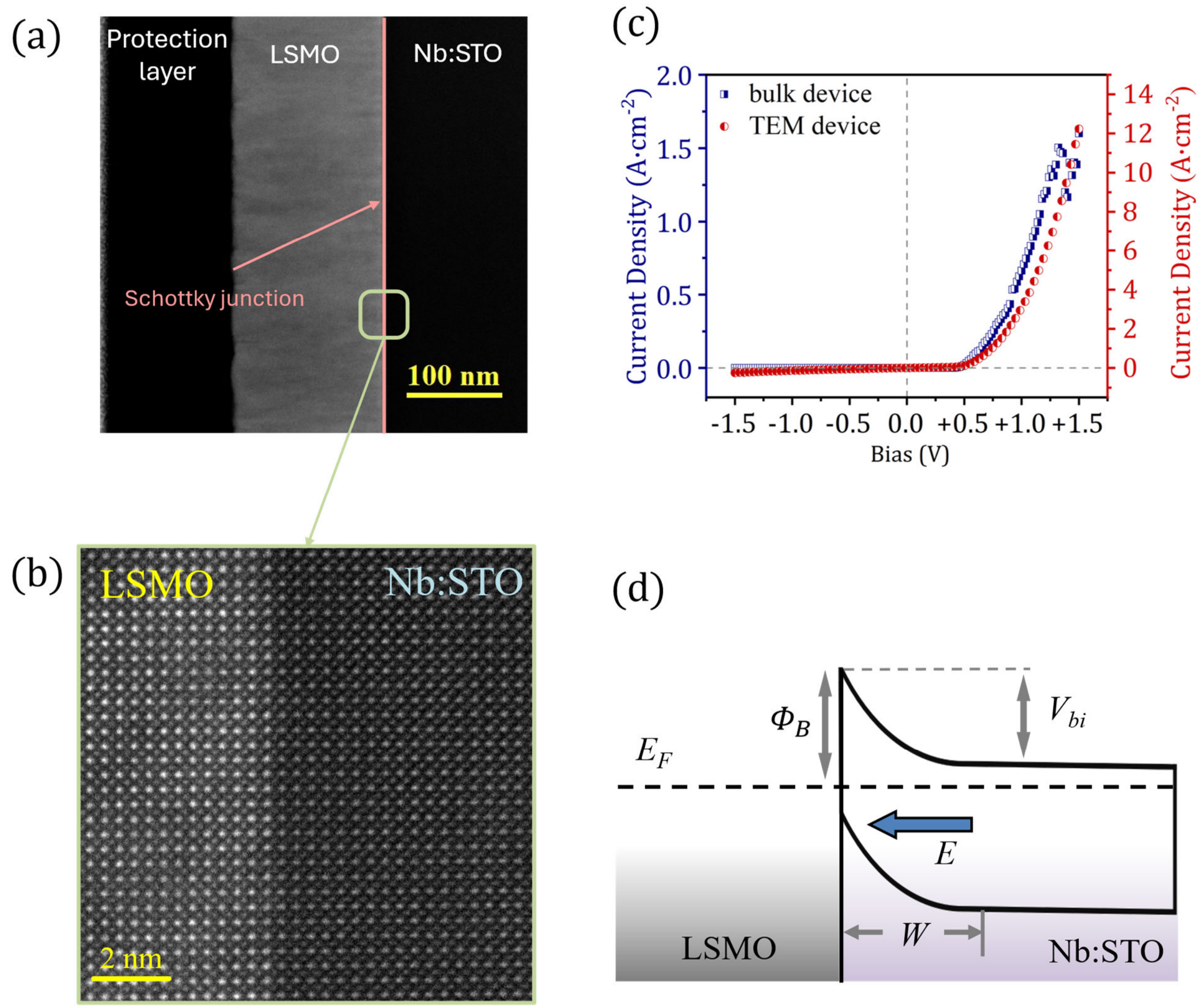

**Figure 2 Configuration of LSMO/STO Schottky junction and electrical validation.** (a) Schematic illustration of the Schottky junction interface and; (b) atomic-resolution ADF-STEM image of the interface; (c) Comparison of $J$-$V$ characteristics for the macroscopic and MEMS devices; (d) schematic equilibrium band diagram illustrating the built-in junction field.

Thus, the macroscopic junction exhibits near-ideal Schottky behavior with $n$ close to 1 and a large parallel resistance ($R_p \approx 30\ G\Omega$) indicating a high-quality junction. The extracted Schottky barrier height of $\varphi_B \approx 0.91$ eV and calculated depletion width of ~63 nm is consistent with values previously reported for LSMO/NSTO Schottky interfaces(*31-33*). The results confirm the excellent quality of the epitaxial oxide junction and establish a reliable foundation for the subsequent *in-situ* electrical measurement.

As might be expected, the MEMS-based TEM device exhibits a less ideal electrical behavior, characterized by an increased ideality factor ($n \approx 2.7$), a higher series resistance ($R_s \approx 3.4$ MΩ) and a significantly reduced parallel resistance ($R_p \approx 290$ MΩ), indicating the presence of additional leakage channels. The increased $R_s$ is most likely attributable to the lamella geometry and the non-ideal electrical contacts formed by FIB-deposited Pt, whereas the reduced $R_p$ likely originates from FIB-induced surface amorphization, material redeposition, and/or Ga implantation, which introduces an electrically inactive surface layer on both sides of the lamella(*24, 34*).

Despite these non-idealities, the extracted transport parameters, together with the pronounced rectifying $J$-$V$ characteristics, demonstrate that the TEM lamella remains an operational Schottky junction and provides a reliable platform for *in situ* electrical characterization. A schematic band diagram of the LSMO/NSTO Schottky junction is presented in fig. 2d, illustrating the built-in electric field within the depletion region. The following sections focus on probing the evolution of this electric field under applied bias at this junction using combined STEM-EBIC and 4D-STEM.

**Table 1 Fitted Schottky parameters from I-V/C-V curves and Cheung-Cheung method.**

| | $n$ | $\varphi_B$ (eV) | $W_0$ (nm) | $R_s$ (MΩ) | $R_p$ (MΩ) |
|---|---|---|---|---|---|
| Bulk device | ~1.1 | ~0.91 | ~63 | - | ~30000 |
| MEMS device | ~2.7 | ~0.67 | ~52 | ~3.4 | ~290 |

**2.2 Operando potential mapping using 4D-STEM**

To quantitatively study the local electric field across the oxide Schottky junction, operando 4D-STEM measurements were performed by tracking the CoM shift of the diffraction pattern. Figs. 3a presents a low magnification overview of the CoM shift in the horizontal ($\mathrm{CoM}_x$) and vertical ($\mathrm{CoM}_y$) directions at zero bias. A magnified BF view from a region across the LSMO/NSTO interface is shown below. The total shift is dominated by the $\mathrm{CoM}_x$ perpendicular to the interface, whereas $\mathrm{CoM}_y$ is negligible (< 10% of $\mathrm{CoM}_x$). Diagonal contrast features are also observed across the specimen. These features originate from diffraction and bend contour contrast, and can produce substantial variations in the diffracted intensities(*25, 26, 35*). To minimize these effects, quantitative measurements near the interface were acquired under different diffraction conditions ('stage precession'). The details of acquisition and calibration procedures are described in the *Materials and Methods* section.

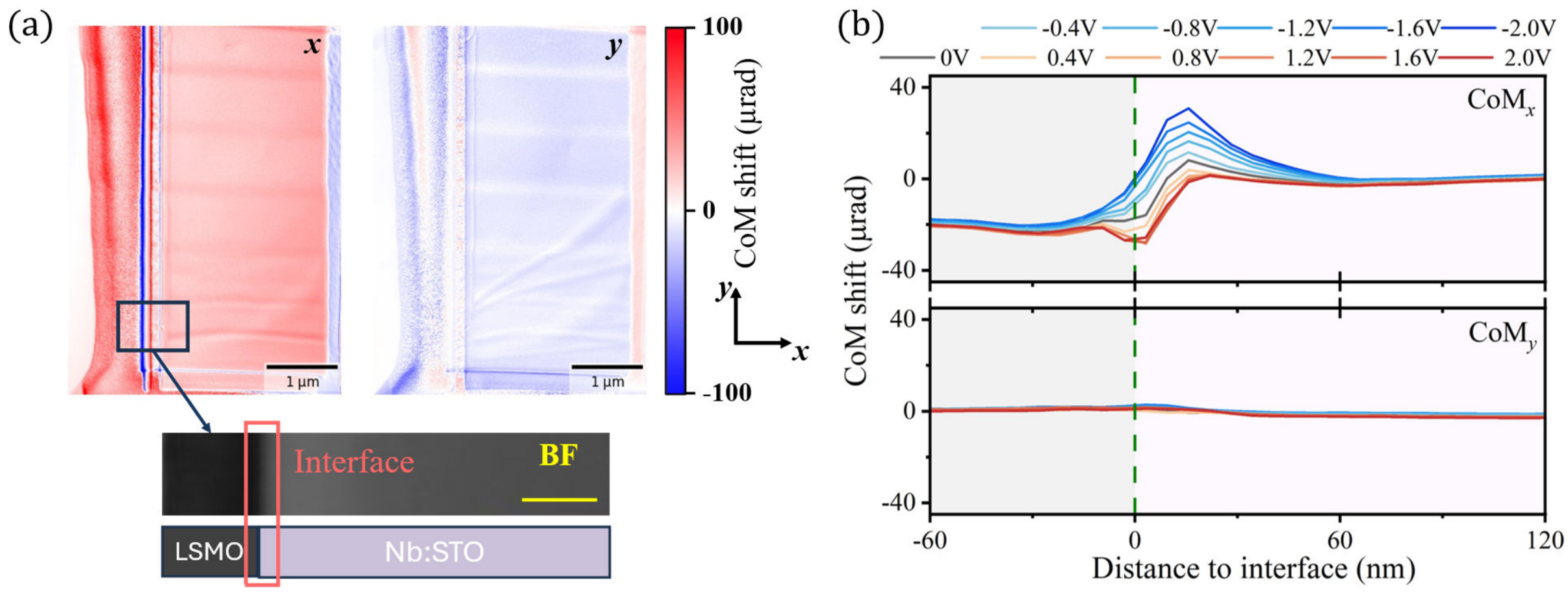

**Figure 3 CoM deflections measurements from 4D-STEM.** (a) CoM along both $x$ and $y$ directions with a large field of view, where a magnified BF image across the interface is shown below, scale bar corresponds to 100 nm; (b) CoM$_x$ (top) and CoM$_y$ (bottom) line profiles at different bias across LSMO/NSTO interface.

Fig. 3b presents CoM$_x$ and CoM$_y$ profiles across the LSMO/NSTO interface for different applied biases. CoM$_y$ remains very small at all biases, while CoM$_x$ exhibits strong bias dependence and is mainly localized around the junction. However, to extract a quantitative measurement of the internal field associated with the Schottky junction from CoM$_x$, several additional steps are required. First, CoM measurements contain contributions not only from the electric field but also MIP and other unwanted contrasts including diffraction contrasts. A change in MIP (ΔMIP) across the heterojunction arises from the differences in crystal structure, density and atomic number of the materials in the device. Previous studies have used additional support from simulations to separate the ΔMIP and electric field contributions in CoM or holography measurements(*25*). Here, taking MIP and other unrelated contrasts to be independent of applied bias we subtract its contribution using a reference state estimating a near-zero-field condition, which can be achieved in a Schottky junction at sufficiently high forward bias(*36*). However, since a significant fraction of applied bias is taken by the large series resistance of the MEMS device at forward bias, establishing this condition cannot be done using 4D-STEM alone. We will demonstrate in section 2.4a how STEM-EBIC measurements can be used to determine near-zero-field conditions and overcome this problem. Second, because the deflection of the electron beam depends only upon the electrically active thickness ($t_{eff}$) of the specimen in which the field is present, an estimate of electrical inactive thickness ($t_{inactive}$) must be estimated and subtracted from the total thickness ($t_{eff} = t_{total} - t_{inactive}$)(*24, 37, 38*). As shown in Figure 1(b) the MEMS device was designed in this work with different thickness steps. From thickness dependent STEM-EBIC measurements, we estimate the $t_{inactive}$ and thus determine the $t_{eff}$ for the quantitative field reconstruction (section 2.4b)(*35, 39*). Finally, any scanning measurement is convoluted with the finite size $d$ of the electron probe, given by(*40*):

$$d \gtrsim \sqrt{\left(\frac{1.22\lambda}{\alpha}\right)^2 + (0.5C_s\alpha^3)^2 + d_g^2 + d_c^2}, \quad (1)$$

where $\lambda$ is the electron wavelength, $\alpha$ is the convergent beam semi-angle (CBSA), $C_s$ is the spherical aberration coefficient, $d_g$ and $d_c$ the source and chromatic aberration limits. In 4D-STEM CoM measurements a small $\alpha$ of ~250 µrad is used to avoid overlap between the direct and diffracted beams, giving a probe size of ~10 nm and producing significant broadening of the measured profile. We describe how this effect can be considered in section 2.4c.

### 2.3 Effects of operando applied bias on STEM-EBIC

STEM-EBIC shows how injected charge carriers respond to the internal electric field, specifically the ability of an internal field to separate electron-hole pairs (see *Materials and Methods*). Fig. 4a shows an ADF image across the interface while the corresponding EBIC signal ($I_{\mathrm{EBIC}}$) at zero bias is shown in fig. 4b. The overlay of both signals, EBIC on the ADF in fig. 4c, allows the electrical response to be directly correlated with local structure. Modest spatial variations of $I_{\mathrm{EBIC}}$ along the direction parallel to the interface are likely associated with local fluctuations in carrier collection efficiency arising from inhomogeneous built-in potential or charge trapping, highlighting the sensitivity of EBIC to variations in the electrical response of the junction at the nanoscale.

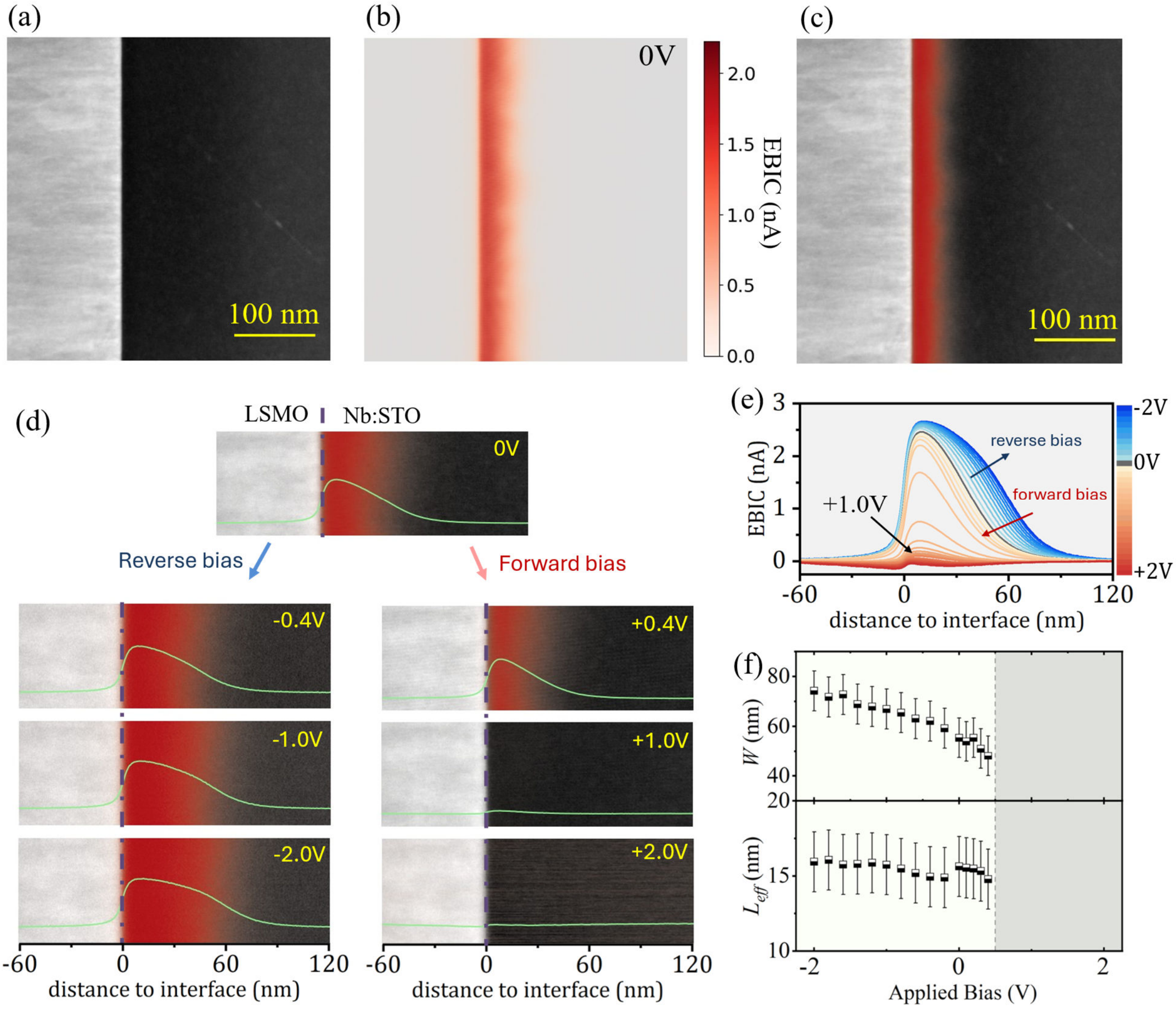

**Figure 4 operando direct probing of electric field evolution using STEM-EBIC.** (a) ADF-STEM image of the junction region; (b) corresponding zero bias EBIC map; (c) overlay of EBIC and ADF signals; (d) bias-dependent STEM-EBIC maps (red) overlaid on ADF-STEM image at different applied biases; (e) Averaged EBIC line profiles extracted across the LSMO/NSTO interface; (f) bias dependence of $L_{eff}$ and $W$, extracted from the EBIC line profiles. (N.B. Analysis is limited to the range over which the EBIC signal remains sufficiently strong for reliable quantification. Fitting procedures are described in *Materials and Methods*.)

Similar overlays for applied bias from −2 V to +2 V are shown in Fig. 4d, with averaged 1D line profiles summarized in fig. 4e. Under reverse bias, the EBIC signal becomes slightly stronger and extends further into the NSTO substrate, consistent with an expansion of the depletion region and a higher junction electric field. Conversely, under forward bias, the EBIC signal progressively weakens and narrows, reflecting the gradual compensation of the built-in electric field. The CBSA for our EBIC measurements is much larger than used for 4D-STEM, ($\alpha$ ~ 30 mrad, giving atomic resolution ADF-STEM images) and convolution with the probe shape is therefore negligible. Rather, resolution is limited by minority carrier movements, i.e. charges injected some distance from the depletion region may still arrive by diffusion and be separated, giving a detectable EBIC signal(*41*).

In a planar junction geometry, as we have in our *n*-type NSTO TEM specimen, the effective hole diffusion length ($L_{eff}$) may by extracted by fitting the long exponential tail of the EBIC profile in the NSTO substrate, i.e. outside the depletion region. Fig. 4f shows that $L_{eff}$ remains constant, within experimental error, at ~16nm for all applied bias. Although substantially smaller than typical diffusion lengths reported for bulk conventional semiconductors, it is consistent with previous EBIC-STEM studies of similar oxide system where the reduced diffusion length is primarily attributed to enhanced surface recombination in thin TEM lamellae(*39, 42, 43*). This relatively short $L_{eff}$ implies reduced broadening of the EBIC signal and is therefore advantageous for resolving fine electric field distribution features with nanometer-scale spatial resolution.

Ong et al. Proposed that the depletion width, $W$, may be determined using the inflection point in the first derivative of $ln(I_{EBIC})$*(41)*, also shown in fig. 4f. There is a clear dependence on applied bias, as expected for a Schottky junction, although the uncertainty in Debye length, and difficulty in inflection point identification at forward bias results in a correspondingly large uncertainty in $W$ (see *Materials and Methods*, and fig. SI10)(*44*). Despite these limitations an estimate of depletion width is helpful to reconstruct the actual electric field distribution, as discussed in section 2.4c below(*44*).

**2.4 Quantitative electric field reconstruction by correlative STEM-EBIC and 4D-STEM**

*a) Purify the electric field contribution from CoM*

The EBIC maps and line profiles in figs. 4d-e show that the EBIC signal is essentially zero for a forward bias of +1V or higher. We therefore infer that above this value the junction field is largely compensated, and further increases in the applied bias only weakly modify the local potential drop across the junction. Under this assumption, $CoM_x$ profiles acquired at +1.2, +1.6, and +2.0 V are averaged to establish a reference state corresponding to a near-zero-field condition, containing all contributions unrelated to the internal electric field, including ΔMIP and any unwanted contrast. The resulting reference profile is shown in the upper right of Fig. 5a, while electric field contributions to the total $CoM_x$ are then obtained by subtracting the near-zero-field reference from the original $CoM_x$ profiles, which are shown below. Here, a single bias-dependent peak is observed at the Schottky interface while the non-interfacial regions in $CoM_x$ are reduced almost to zero, indicating that contributions unrelated to the internal field have been removed.

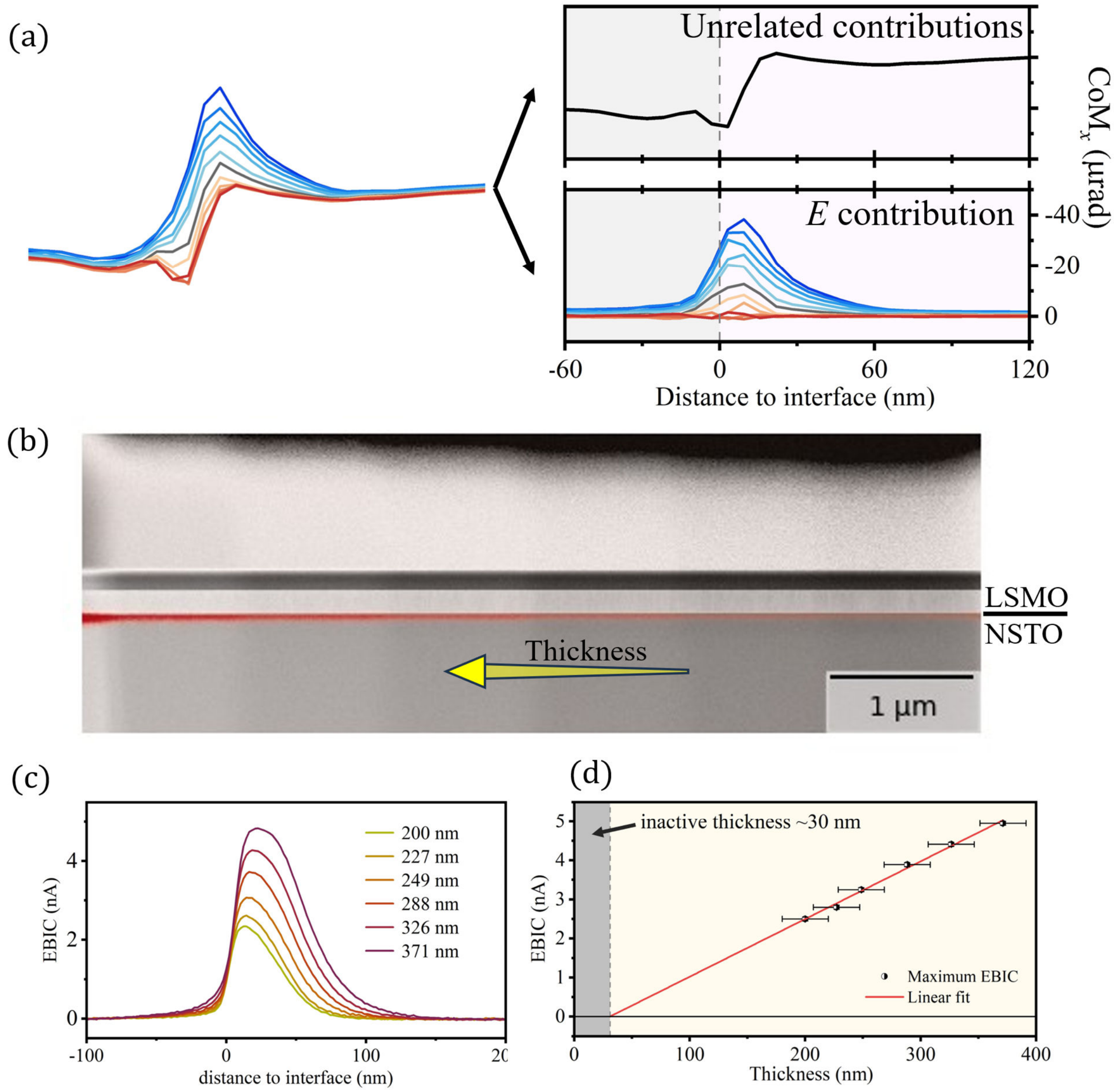

**Figure 5 Separation of electric field contributions and effective thickness calibration.** (a) separation of electric field contribution (right bottom) from $CoM_x$ (left) in fig. 3b using a near-zero-field condition as reference (right top); (b) EBIC overlaid on ADF-STEM image across different thickness steps. (c) EBIC line profiles across the interface at different thickness; (d) max EBIC signal against EELS measured thickness used to determine the effective thickness.

*b) Determining effective thickness $t_{eff}$*

Fig. 5b shows the zero-bias EBIC-ADF overlaid image across the stepped thickness region of the lamella, while the total thickness $t_{total}$ for each step is measured by EELS, as described in *Materials and Methods* and fig. SI6. The number of electron-hole pairs generated by the electron beam increases with the electrically active volume of the specimen(*35*), i.e. proportional to $t_{eff}$. Therefore, the maximum signal is expected to increase with $t_{eff}$, as can be seen in the EBIC line profiles at different $t_{total}$ (fig. 5c). A plot of maximum $I_{EBIC}$ against $t_{total}$ is shown in fig. 5d, where an approximately linear dependence is observed over the investigated thickness range. Extrapolation of this relationship to zero EBIC signal gives a $t_{inactive}$ approximately 30 ± 5 nm. Thus, for any given thickness the $t_{eff}$ used for the quantitative electric field calculation is therefore obtained by subtracting $t_{inactive}$ from $t_{total}$.

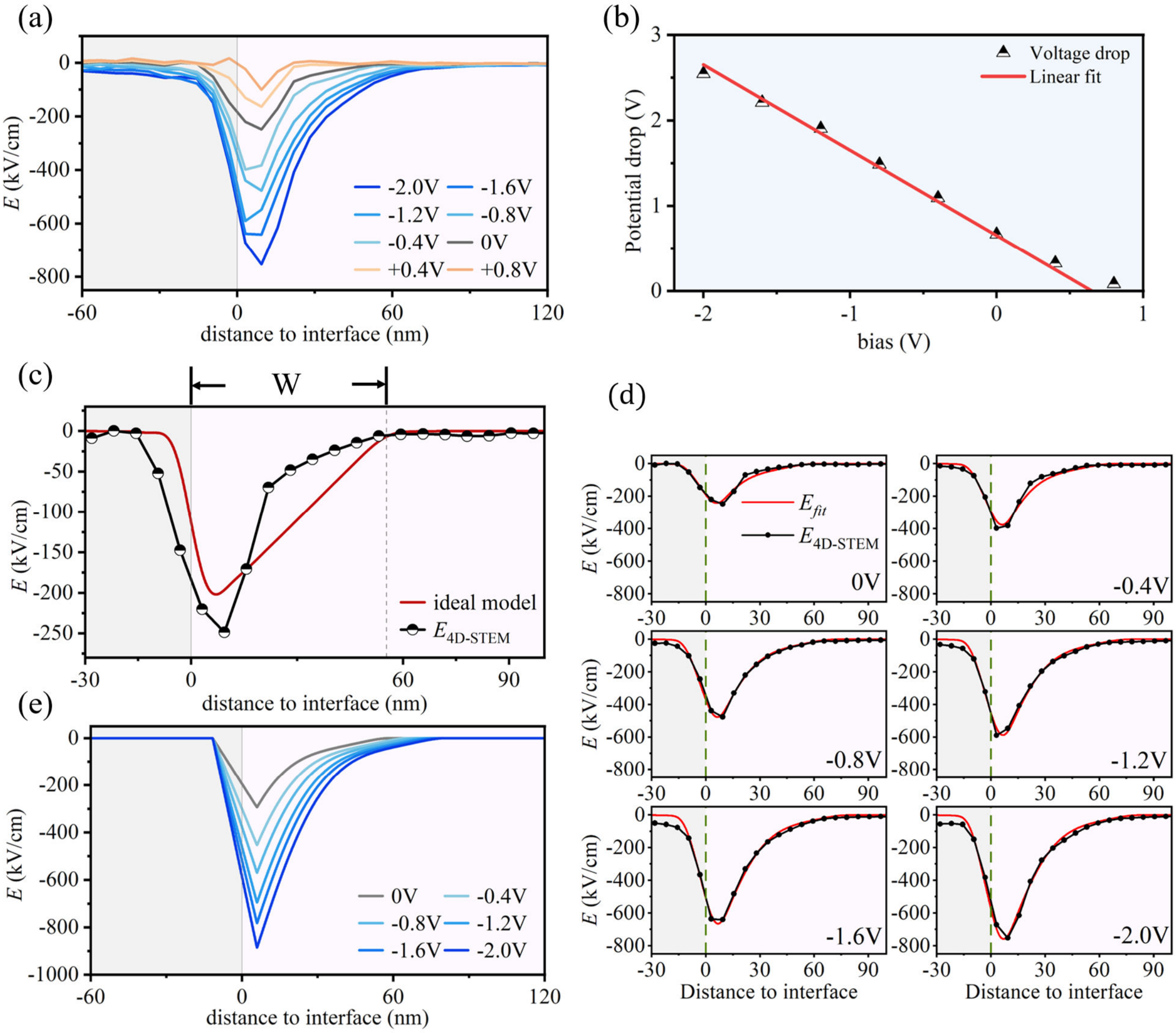

**Figure 6 Quantitative electric field reconstruction.** (a) calculated electric field distributions at different bias from separated $CoM_x$ using $t_{eff}$; (b) integrated junction potential drop against applied

bias; (c) comparison between measured electric field and a convolved ideal Schottky model at zero bias; (d) fitting from forward modelling at different bias; (e) reconstructed original field distribution.

*c) Quantitative measure of internal electric field and the effect of finite probe size*

Using the field-related contributions of $\mathrm{CoM}_x$ and $t_{eff}$, the experimentally observed electric field distribution $E_{\mathrm{obs}}(x)$ may be calculated as shown in Fig. 6a (see *Materials and Methods*). Integrating the line profiles gives the potential drop across the interface, shown in Fig. 6b, where a reliable linear fit can be obtained using a fixed slope of -1 in reverse bias (N.B. in reverse bias the junction resistance is much larger than any series resistance, and so the latter may be neglected). The built-in potential $V_{bi}$ at zero bias is estimated to be ~0.65 ± 0.04 V, which is in good agreement with $\varphi_B \approx 0.67$ eV from the $J$-$V$ analysis. Under high forward bias when the junction is open, the series resistance takes a larger fraction of the applied voltage, so the voltage across the junction is smaller (0.03V at +0.8V).

Interestingly, the internal field $E_{\mathrm{obs}}(x)$ in Fig. 6a is significantly different from an ideal metal/*n*-type semiconductor Schottky junction with uniform doping. In a 1D case under Poisson's depletion approximation and estimating the net space charge density ($\rho$) equal to donors that fully ionized, the gradient of the internal electric field is given by(*44*):

$$\frac{dE}{dx} = \frac{\rho(x)}{\varepsilon_0 \varepsilon_r} = \frac{qN_D}{\varepsilon_0 \varepsilon_r}, \tag{2}$$

where $q$ is the elementary charge, $N_D$ is the donor concentration, $\varepsilon_r$ is the relative permittivity, $\varepsilon_0$ is the vacuum permittivity and $x$ is the distance from the interface. With perfect screening in the metal, constant permittivity and uniform doping, the electric field is then:

$$E(x) = \begin{cases} 0 & x \leq 0 \; and \; x \geq W \\ -\frac{qN_D}{\varepsilon_0 \varepsilon_r}(W - x) & 0 < x < W \end{cases}, \tag{3}$$

i.e. the field changes abruptly at the interface and decreases linearly to zero over the depletion region. However, $E_{\mathrm{obs}}(x)$ in fig. 6a has a gradual rise at the LSMO side, and the decay within the NSTO depletion region is clearly not linear. As noted above, one reason for this, could be the finite electron probe size that broadens the measured field(*27*). Thus, we convolute this ideal model with a Gaussian electron probe shape ($K$, Fig. SI7) and compare with $E_{\mathrm{obs}}$ at zero bias (using $W_0$ = 57 nm from STEM-EBIC and $V_{bi}$ = 0.65V), as shown in fig. 6(c). Although the probe convolution smooths the field rise on the metal side, the resulting profile still fails to reproduce the nonlinear decay of $E_{\mathrm{obs}}$ in NSTO or the full extent of its extended penetration into the LSMO.

The above suggests that a more sophisticated model is required for this oxide Schottky junction. The extended electric field penetration into LSMO may be explained by a reduced (metallic) density of states in the first few unit cells(*45*). Such interfacial electronic degradation, commonly described as an electrically 'dead' layer, has been proposed in LSMO/STO heterostructures(*45-47*).

Within the depletion region, as the local slope of the electric field is proportional to $\rho/\varepsilon_r$ in Eq. 2, the nonlinear decay of the electric field suggests a variation in permittivity and/or space charge density. One explanation is the field dependent in $\varepsilon_r$ of STO(*48*). STO possess a much larger relative permittivity $\varepsilon_r$ than conventional semiconductors such as silicon or gallium arsenide (where $\varepsilon_r$ can be taken as constant as in an ideal semiconductor model(*44*)). Several workers have suggested a decrease in $\varepsilon_r$ at high electric fields(*48-50*). Incorporating a finite interfacial region at the interface, which is estimated to be a linear decay for electric fields, together with a field dependent $\varepsilon_r$. For simplicity here we take $N_D$ ($\rho$) to be constant, and the total electric field distribution may then be expressed as:

$$E(x,V) = \begin{cases} f(W-x) & d_{NSTO} \leq x \leq W(V) \\ E(d_{NSTO},V)\frac{x-d_{LSMO}}{d_{NSTO}-d_{LSMO}} & d_{LSMO} \leq x \leq d_{NSTO} \\ 0 & x \leq d_{LSMO} \text{ and } x \geq W(V) \end{cases}, \quad (4)$$

where $d_{LSMO}$ and $d_{NSTO}$ define the interfacial layer for both LSMO and NSTO, and $f(W - x)$ is a polynomial function sharing at different bias. Using $W(V)$ measured from STEM-EBIC at zero and reverse bias, the corresponding original field distribution can be reliably fitted using eq. 4. The detailed fitting procedures are described in *Materials and Methods* section. As shown in fig. 6d, the convolved field model reproduced $E_{\mathrm{obs}}$ at zero and negative bias and the fitted original field distribution is shown in fig. 6e. These results demonstrate that a field dependent permittivity and a finite interfacial region can well describe the current oxide Schottky heterojunction, even though this model simplifies the transition point, the interfacial layer and metal screening, where the reconstructed field distribution may not be a unique solution.

The nonlinear decay of electric field profiles in the depletion region is consistent with previous studies deduced from indirect measurements, where STO-based junctions consistently exhibit anomalous $C$-$V$ characteristics(*48, 51*). However, the variation in $\rho$ cannot be ruled out, which is another possibility causing the nonlinear decay. This may arise from non-uniform doping concentration next to the junction, or compensation of doping either by changes in charge state and/or the presence of charged point defects(*52*). Thus, while atomic resolution STEM indicates a structurally commensurate interface that is free of defects such as dislocations, there are several possible mechanisms, that could be in play simultaneously, to give a nonlinear field decay.

In summary, these results provide the first nanometer scale reconstruction of the bias-dependent electric field evolution within the depletion region of an operating oxide Schottky junction, demonstrating a solid approach using correlative STEM-EBIC and 4D-STEM to measure the electric field distribution. The reconstructed electric field distribution profiles also provide direct spatial resolved demonstration of the finite interfacial layer and nonlinear electric field distribution within the depletion region, highlighting the characteristics of STO-based Schottky junctions that can only be deduced from indirect electrical or photoemission measurements(*48, 49*).

## 3. Conclusion

In conclusion, a spatially resolved electric field distribution at nanometer scale is reconstructed, for the first time, in an oxide Schottky heterojunction combining 4D-STEM and STEM-EBIC. We successfully integrated LSMO/NSTO heterojunction into MEMS device which is verified to be operational and preserves rectifying behavior. Combining with STEM-EBIC, we overcame several key challenges in 4D-STEM towards heterojunctions including inactive thickness, and identification of the near-zero-field condition for any unrelated contrasts including ΔMIP. STEM-EBIC also provides important constraints of the depletion width so the original field distribution can be reconstructed using forward modelling. The reconstructed field distributions show nonlinear decay within the depletion region of NSTO with an extended field penetration into the LSMO, showing good agreement with the indirect measurements from previous studies.

Our results demonstrate that STEM-EBIC can be successfully implemented in oxide heterostructures and provide complementary information that is particularly useful to overcome the limitation of 4D-STEM. More broadly, our results highlight the opportunities for conducting nanoscale electrical characterization in oxides materials within TEM. Similar device configurations can also be realized in other systems, including those on insulating substrates, as demonstrated by the rectifying $I$-$V$ characteristics of a Pt/STO junction shown in fig. SI8. Beyond demonstrating the effectiveness of STEM-EBIC in STO-based oxide systems, our results suggest that the technique may be particularly well suited to oxides because of their short diffusion lengths. The implementation of STEM-EBIC may therefore help overcome some of the current limitations of 4D-STEM in oxides and enable the characterization of finer electric field structures at nanometer scale, such as ferroelectric domains and 2-dimensional electron gas.

## 4. MATERIALS AND METHODS

### 4.1 Sample Growth and Macroscopic Device Fabrication

Epitaxial $La_{0.67}Sr_{0.33}MnO_3$ thin films were grown on commercially available 0.05 wt.% Nb-doped $SrTiO_3$ (001) single crystal substrate by pulsed laser deposition (PLD). A KrF excimer laser (λ= 248 nm) was operated at a repetition rate of 3 Hz with a laser fluence of approximately 1.5 $J \cdot cm^{-2}$. During deposition, the substrate temperature was maintained at 650 °C under an oxygen pressure of 0.3 mbar. After deposition, the samples were cooled to room temperature at 10 $°C \cdot min^{-1}$ under the same oxygen pressure. The nominal donor concentration of the NSTO substrate was estimated to approximately $1.6 \times 10^{25}$ $m^{-3}$.

Macroscopic devices on LSMO/NSTO were fabricated using photolithography designed the top contact area ( 285 µm × 285 µm ), where a probe was in direct contact. An InGa electrode was applied to the backside of the NSTO substrate serving as the other ohmic contact.

### 4.2 TEM device fabrication

The MEMS device was prepared from the macroscopic LSMO/NSTO sample using a TESCAN AMBER Ga FIB-SEM. A Pt protection layer was first deposited by electron-beam on the LSMO to minimize surface damage, followed by ion-beam deposited Pt layer. An additional $SiO_2$ insulating layer was then deposited on top of the Pt layers to reduce conductive redeposition and minimize the possibility of leakage or short-circuiting during subsequent thinning process.

Cross-sectional lamellae were then prepared by standard FIB lift-out procedures along <100> direction and transferred onto a MEMS-based biasing chip (DENSsolutions), as shown in fig. 1b. The lamella was then connected to the MEMS electrodes by Pt deposition on both sides to establish electrical contact for biasing TEM measurements. Additional isolation cuts were made to define the final circuit geometry.

Initial thinning was performed at 30 kV using beam currents between 50 and150 pA. The thickness steps used for the electrically active thickness calibration were produced during the final shaping at 20 kV and 40 pA. The sample was then polished at 5kV and finally at 2 kV to reduce FIB induced surface damage. The final designed circuit geometry is shown in fig. SI1, while additional compositional and interface characterization is presented in Fig SI2 and SI3 respectively.

**4.3 Electrical measurement**

Current-voltage ($I$-$V$) curve of the macroscopic LSMO/NSTO devices were measured at room temperature using a Keithley 2636B System SourceMeter over a bias range from −2 to +2 V in a hysteresis loop. *C-V* measurements were performed using an Agilent E4980A Precision LCR Meter. The $I$-$V$ curve of MEMS device was measured between -2.5V and +2.5V using a point electronic GmbH Electrical Analysis system. The parallel resistance, $R_p$, was estimated from the inverse slope of the approximately linear $I$-$V$ response near-zero-bias, between -0.05 V and 0.05 V. The $I$-$V$ characteristics were analyzed using the thermionic emission model given by(*53*):

$$I = A^{*}A_{eff}T^{2}exp\left(-\frac{q\varphi_B}{k_BT}\right)\left[exp\left(\frac{qV}{nk_BT}\right)-1\right], \tag{5}$$

where $q$ is the charge of the electron, $k_B$ is the Boltzmann constant, $A_{eff}$ is the effective area, $T$ is the temperature, $A*$ is the Richardson constant (156 $A\cdot cm^{-2}\cdot K^{-2}$ from intrinsic STO)(*54*), $\varphi_B$ is the zero bias Schottky barrier height, and $n$ is the ideality factor. The effective area $A_{eff}$ of the MEMS device was estimated to be approximately 3.2 $\mu m^2$ using SEM measurement after accounting for the electrically inactive thickness (fig. SI9). A plot of $1/C^2$ against $V$ (fig. SI4) gives:

$$\frac{1}{C^2} = \frac{2(V_{bi}-V)}{qA^2\varepsilon_0\varepsilon_r N_D}, \tag{6}$$

where built-in potential $V_{bi}$ is obtained from the intercept of the linear fit. After calibrating the current density $J = I\cdot A_{eff}^{-1}$, $n$ and $\varphi_B$ can be directly extracted from linear fitting from the $J$-$V$ characteristics for the macroscopic device neglecting $R_s$.

For the MEMS device, Cheung-Cheung method is employed to determine $n$, $R_s$, and $\varphi_B$(*30*):

$$\frac{d(V)}{d(\ln J)} = R_sA_{eff}J + \frac{nk_BT}{q}, \tag{7}$$

and:

$$H(J) \equiv V - \left(\frac{nk_BT}{q}\right)\ln\left(\frac{J}{A^*T^2}\right) = R_sA_{eff}J + n\varphi_B. \quad (8)$$

Linear fitting of $dV/d(lnJ)$ against $J$ gives $R_s$ and $n$, while linear fitting of $H(J)$ against $J$ gives $\varphi_B$ and another estimation of $R_s$. The fitting ranges and the corresponding uncertainties are shown in fig. SI5.

### 4.4 STEM imaging and spectroscopy

STEM experiments were performed in a double aberration-corrected JEOL ARM200F operated at 200kV and equipped with DENSsolutions lightening holder. ADF-STEM images were acquired using a convergence semi-angle of 23.5 mrad, and a detector collection range from 60-180 mrad. EDS measurements were performed using an Oxford Instruments windowless energy-dispersive X-ray detector. EELS measurements were acquired at the same condition using a collection semi-angle below 30mrad.

### 4.5 4D-STEM

4D-STEM datasets were acquired using a Quantum Detectors Merlin detector operated at 2 × 6/12- bit depth. A convergence semi-angle of approximately 250 µrad was used in low magnification Lorentz mode. Operando 4D-STEM measurements were performed using four-point stage-precession (±1°) at applied biases between −2 and +2 V in steps of 0.4 V. The specimen was tilted approximately 6-8°away from the zone-axis in the horizonal direction relative to the junction and by less than 1° in the vertical direction. Equivalent vacuum datasets were acquired for each precession condition and used to correct beam shifts, specimen tilt, and instrumental offsets. The datasets were registered in both scan and diffraction plane before averaging.

The electric field was determined from the CoM shift of the direct beam on the detector when scanning at a point with an electric field using the following equation derived from Ehrenfest's Theorem(*55, 56*):

$$\vec{E}_\perp = -h \cdot \frac{\sin(\theta)}{\lambda} \cdot \frac{v}{t_{eff} \cdot e}\vec{r}, \quad (9)$$

where $E$ is the projected electric field, $h$ is Planck's constant, $\theta$ is the beam deflection angle, which is calculated from the CoM shift, $\lambda$ is the wavelength, $e$ is the charge, $v$ is the relativistic velocity of the electron.

The effective thickness of the material is calibrated using the absolute thickness obtained from EELS in NSTO close to the junction using Digital Micrograph, as described in fig. SI6. This is using the mean free path length of intrinsic STO, where Nb doping is neglected. The measured thicknesses are 200 ± 20 nm, 227 ± 23nm, 249 ± 25 nm, 288 ± 29 nm, 326 ± 33 nm and 371 ± 37 nm.

### 4.6 STEM-EBIC

STEM-EBIC measurements were performed using the point electronic GmbH Electrical Analysis system. The top contact of the lamella was connected to a transimpedance amplifier, followed by a secondary voltage amplifier to extract the EBIC signal, while the bottom contact was connected to a voltage source. The same baseline correction procedure was applied to the two-dimensional EBIC maps acquired at each bias. One-dimensional profiles were then obtained by integrating the signal parallel to the interface to give the EBIC intensity as a function of distance perpendicular to the junction.

The depletion width, $W$, was estimated from the one-dimensional EBIC profiles at different bias using the inflection points of $dln(I_{EBIC})/dx$ following reference(*41*):

$$W = (x_r - x_l) - 2r_{K_{EBIC}}, \tag{10}$$

Where $x_l$ is the first inflection position and $x_r$ is the second inflection position in the derivative plots, $r_{K_{EBIC}}$ is the radius of the EBIC probe. Before differentiation, the EBIC profiles were smoothed using Savitzky-Golay filter at a window of 41 points (for 373 total data points) to reduce high-frequency noise. The same smoothing procedure was applied to all bias conditions. Representative $dln(I_{EBIC})/dx$ profiles are shown in Fig. SI10. The left inflection point remains approximately at the same position with applied bias, while the right inflection point progressively shifts, reflecting the change in depletion width. Under forward bias, however, the EBIC signal becomes weaker, and the right inflection point is significantly broadened, making its position more difficult to determine accurately. For this reason, the depletion width obtained under these conditions has a larger uncertainty, and values for which the inflection point cannot be reliably identified are not used as quantitative constraints in the electric field reconstruction.

Outside the depletion region, minority carriers generated in the neutral region in NSTO can diffuse towards the junction and contribute to the measured EBIC signal. Under low injection regime, this contribution can be described by the probability that an electron–hole pair reaches the edge of the electric field region, given by(*41, 42*):

$$I_{EBIC} = k(x_L)^{\alpha} e^{\left(-\frac{x_L}{L_{eff}}\right)}, \tag{11}$$

where $k$ is a constant, $x_L$ is the distance from the edge of depletion region, $\alpha$ is the linearization parameter. The effective hole diffusion, $L_{eff}$, was obtained by nonlinear least-squares (NLLS) of the one-dimensional EBIC profile tail. The same fitting procedure was applied to all bias conditions for which a sufficiently strong EBIC signal was obtained. Representative fits are shown in fig. SI11.

### 4.7 Forward modelling procedures

The electric field profiles extracted from 4D-STEM were analyzed using a forward model described by eq. 4, and $f$ is given by:

$$f(z) = az + bz^2 + cz^3 + \cdots, \tag{12}$$

where $z = W - x$. While the parameters defining $f$ and the positions of the interfacial transition regions $d_{\mathrm{LSMO}}$ and $d_{\mathrm{NSTO}}$, were shared across all bias conditions with $-20$ nm$< d_{\mathrm{LSMO}} < 0$, and $0 < d_{\mathrm{NSTO}} < 10$ nm. $W(V)$ was allowed to vary with applied bias within the uncertainty determined independently from the STEM-EBIC measurements.

The model was fitted simultaneously to the electric field profiles acquired at zero bias and under reverse bias. Before comparing the experimental data, the calculated field distribution for each bias condition was convolved with the experimental 4D-STEM probe function $K$ (fig. SI7). The model parameters were obtained by NLLS optimization of the global mean squared error (MSE) between the probe convolved model and 4D-STEM measured electric field over the same spatial range, -30 nm < $x$ < 100 nm, for all bias conditions. The polynomial order of $f$ was progressively increased from first other, and the corresponding global MSE was evaluated after each fit. The MSE decreased with increasing polynomial order and showed non-substantial improvement beyond fifth order, which is used for the final reconstruction, as shown in fig. SI12.

## 6. ACKNOWLEDGEMENT

We thank Warwick RTP for the use of instruments. This work is supported by the Engineering and Physical Sciences Research Council (Grant No. EP/V028596/1). This work is also supported by the Quantum Science and Technology-National Science and Technology Major Project (Grant No. 2024ZD0300104), the National Key Research and Development Program (Grant No. 2025YFE0201100) and Anhui Provincial Natural Science Foundation (Grant No.2508085ZD015). Yining Xie would like to acknowledge China Scholarship Council (CSC) for funding.

## 7. Contribution

***Ana S*** and ***Y. X*** conceived the idea with the discussion with ***M. Y***. ***X. L*** grew the LSMO/NSTO macroscopic device and conducted electrical measurements under the supervision of ***M. Y***. ***Aditya S*** and ***M. A*** prepared the Pt/STO macroscopic device. ***Y. X*** fabricated the MEMS devices and conducted the STEM experiments with ***E. M*** under the supervision of ***Ana S** and **R. B***. ***Y. X*** wrote the manuscript with input from all authors.

## 8. Declaration

The authors declare that they have no competing interests.

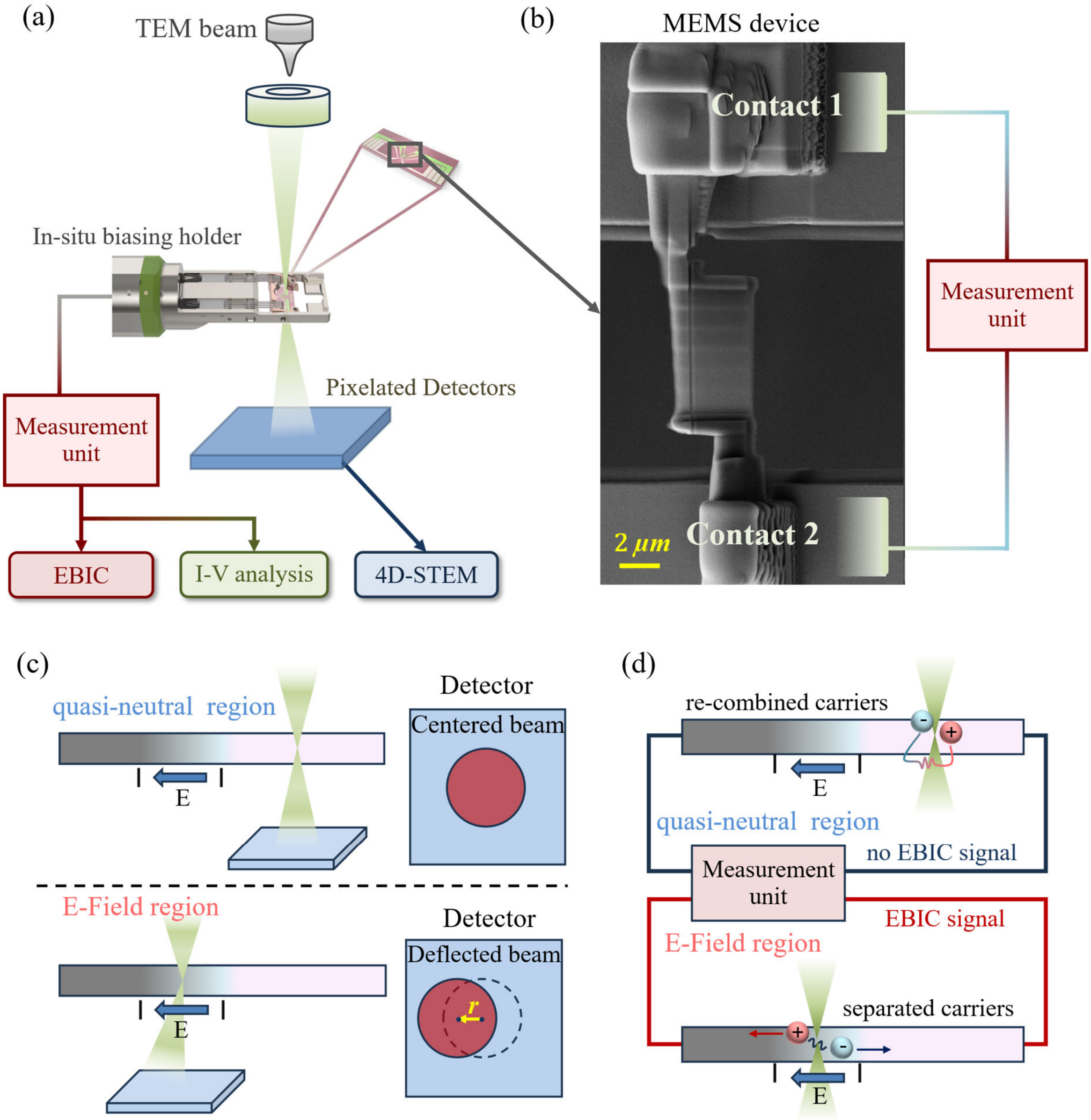

**Figure 1 In-situ experimental configuration.** (a) Schematic of the in-situ STEM setup, where pixelated detectors acquire 4D-STEM datasets, while a biasing holder applies external bias and collects EBIC signal; (b) SEM image of the device geometry; (c) schematic diagram of 4D-STEM measuring internal electric field from a quasi-neutral region without an electric field (top) and under electric field presenting (bottom); (d) schematic diagram of STEM-EBIC from a quasi-neutral region (top) and under electric field presenting (bottom).

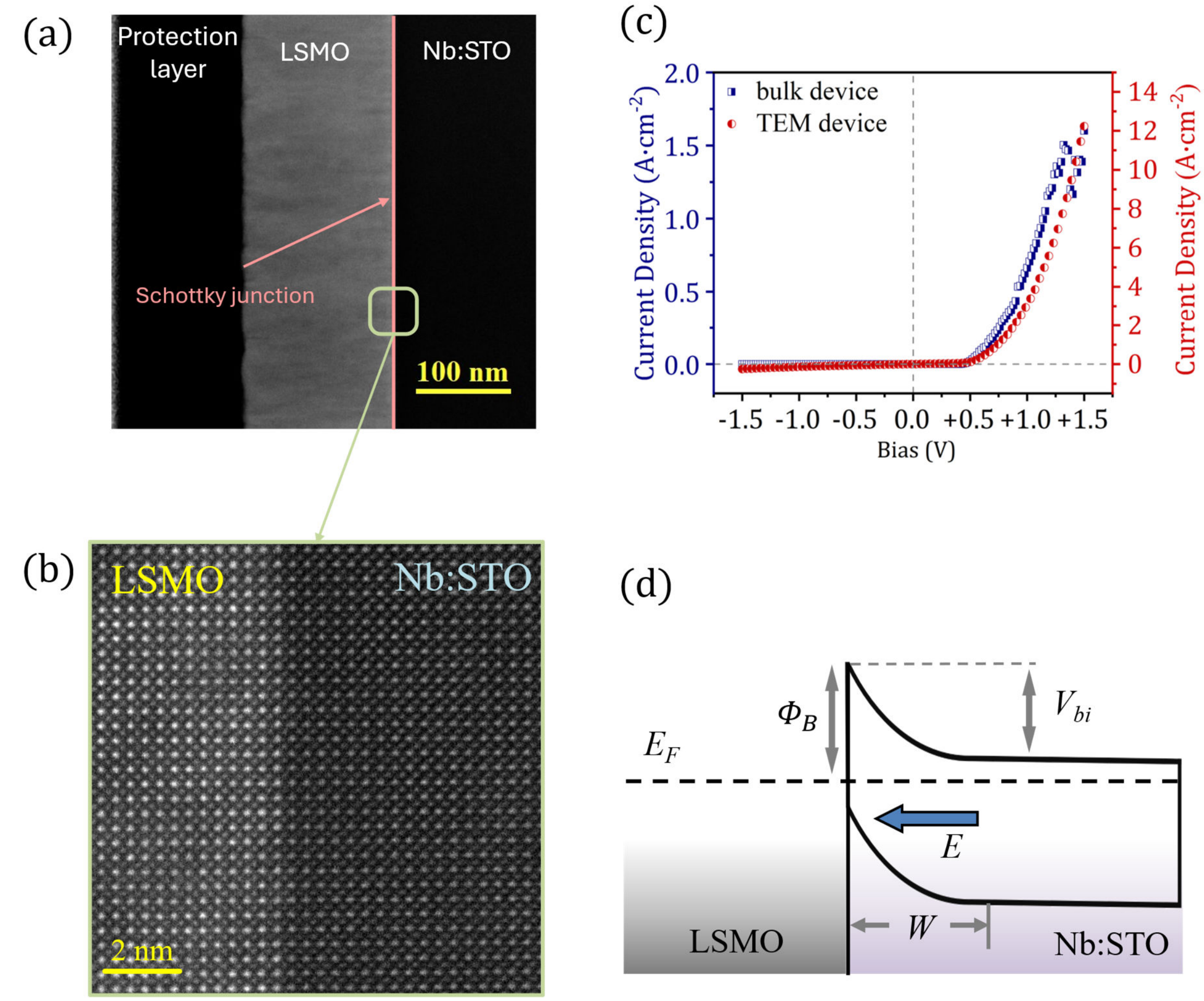

**Figure 2 Configuration of LSMO/STO Schottky junction and electrical validation.** (a) Schematic illustration of the Schottky junction interface and; (b) atomic-resolution ADF-STEM image of the interface; (c) Comparison of $J$-$V$ characteristics for the macroscopic and MEMS devices; (d) schematic equilibrium band diagram illustrating the built-in junction field.

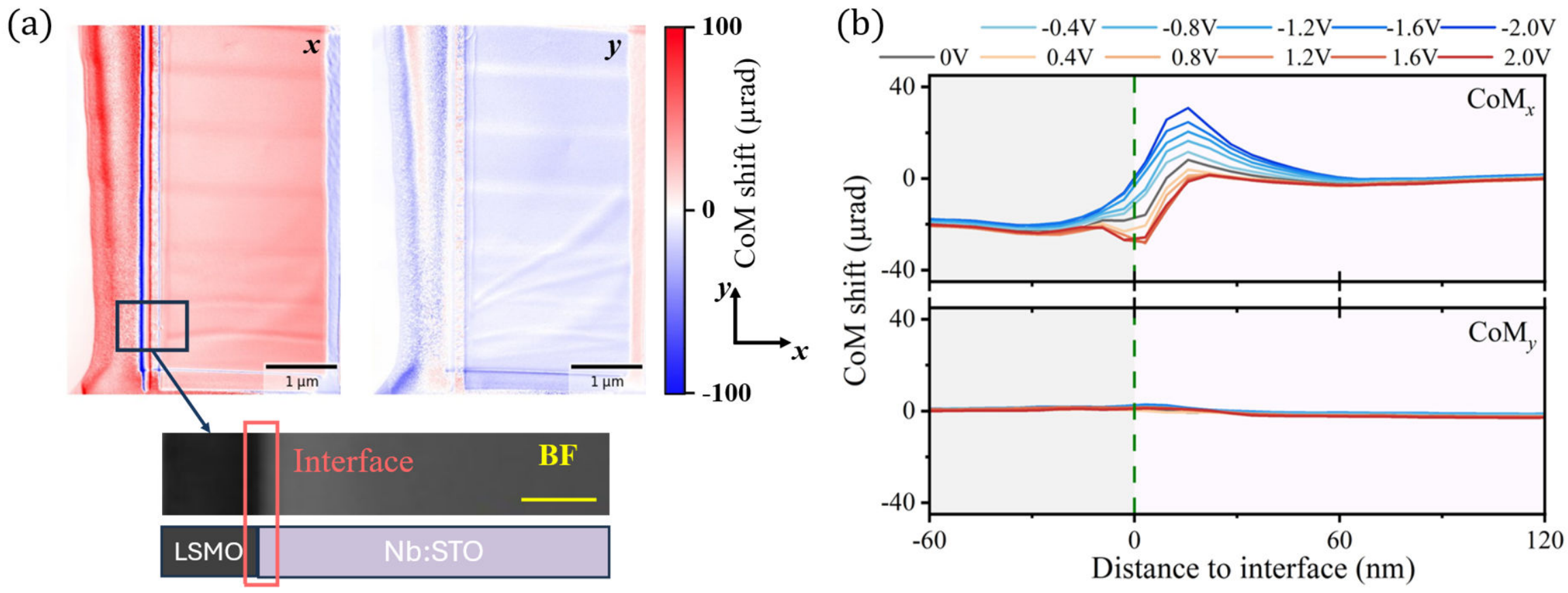

**Figure 3 CoM deflections measurements from 4D-STEM.** (a) CoM along both $x$ and $y$ directions with a large field of view, where a magnified BF image across the interface is shown below, scale bar corresponds to 100 nm; (b) $\mathrm{CoM}_x$ (top) and $\mathrm{CoM}_y$ (bottom) line profiles at different bias across LSMO/NSTO interface.

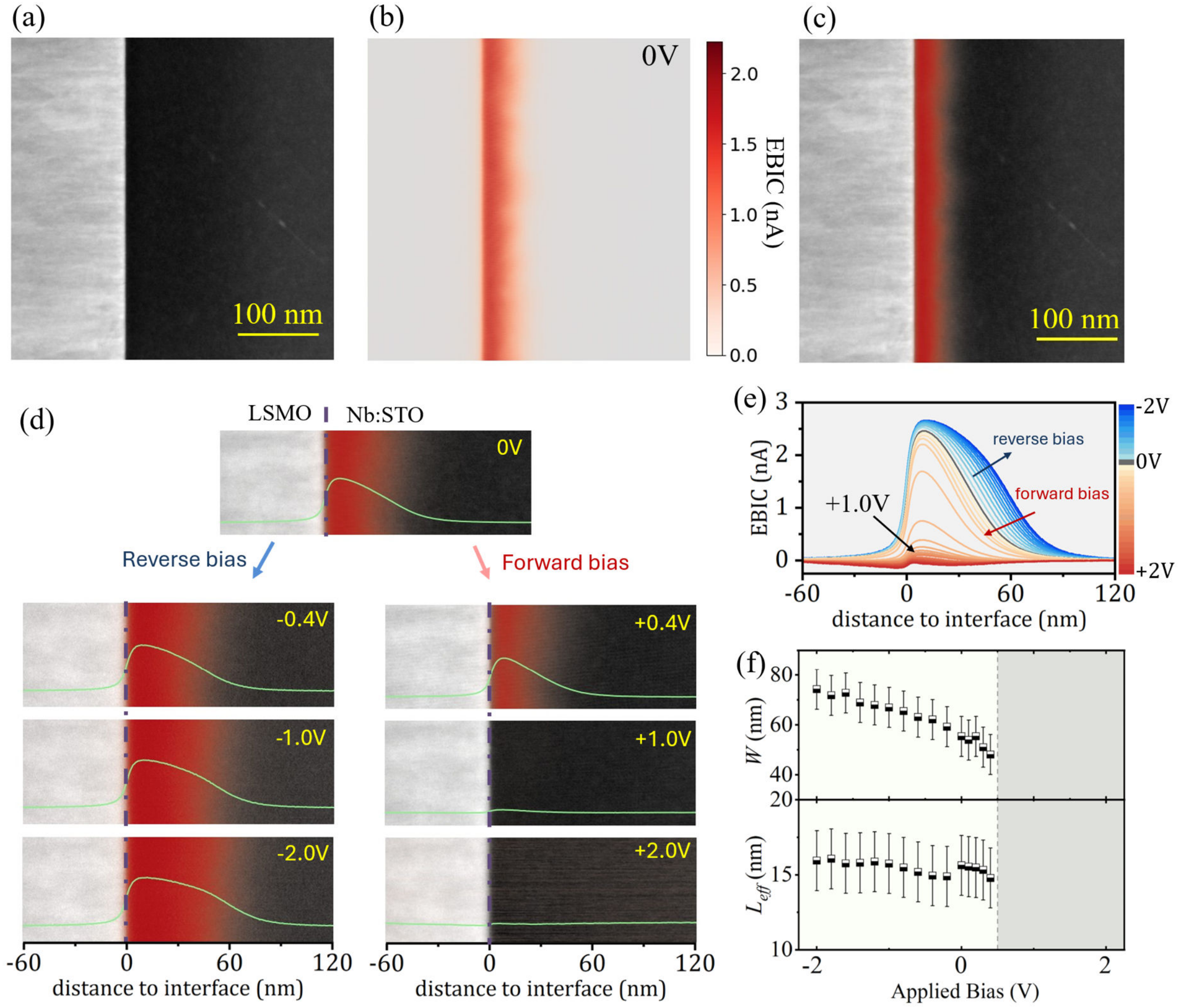

**Figure 4 operando direct probing of electric field evolution using STEM-EBIC.** (a) ADF-STEM image of the junction region; (b) corresponding zero bias EBIC map; (c) overlay of EBIC and ADF signals; (d) bias-dependent STEM-EBIC maps (red) overlaid on ADF-STEM image at different applied biases; (e) Averaged EBIC line profiles extracted across the LSMO/NSTO interface; (f) bias dependence of $L_{eff}$ and $W$, extracted from the EBIC line profiles. (N.B. Analysis is limited to the range over which the EBIC signal remains sufficiently strong for reliable quantification. Fitting procedures are described in *Materials and Methods*.)

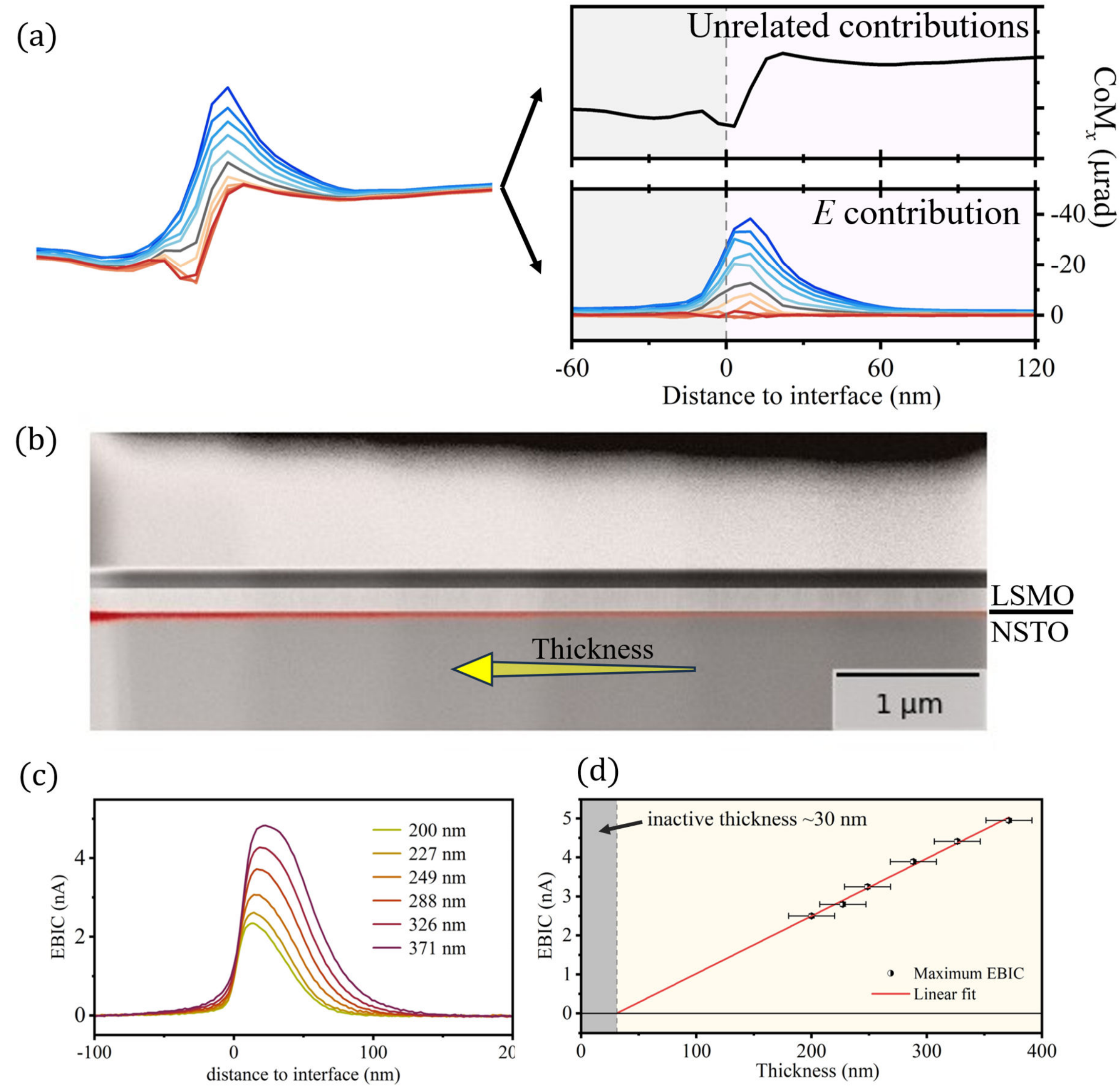

**Figure 5 Separation of electric field contributions and effective thickness calibration.** (a) separation of electric field contribution (right bottom) from CoM$_x$ (left) in fig. 3b using a near-zero-field condition as reference (right top); (b) EBIC overlaid on ADF-STEM image across different thickness steps. (c) EBIC line profiles across the interface at different thickness; (d) max EBIC signal against EELS measured thickness used to determine the effective thickness.

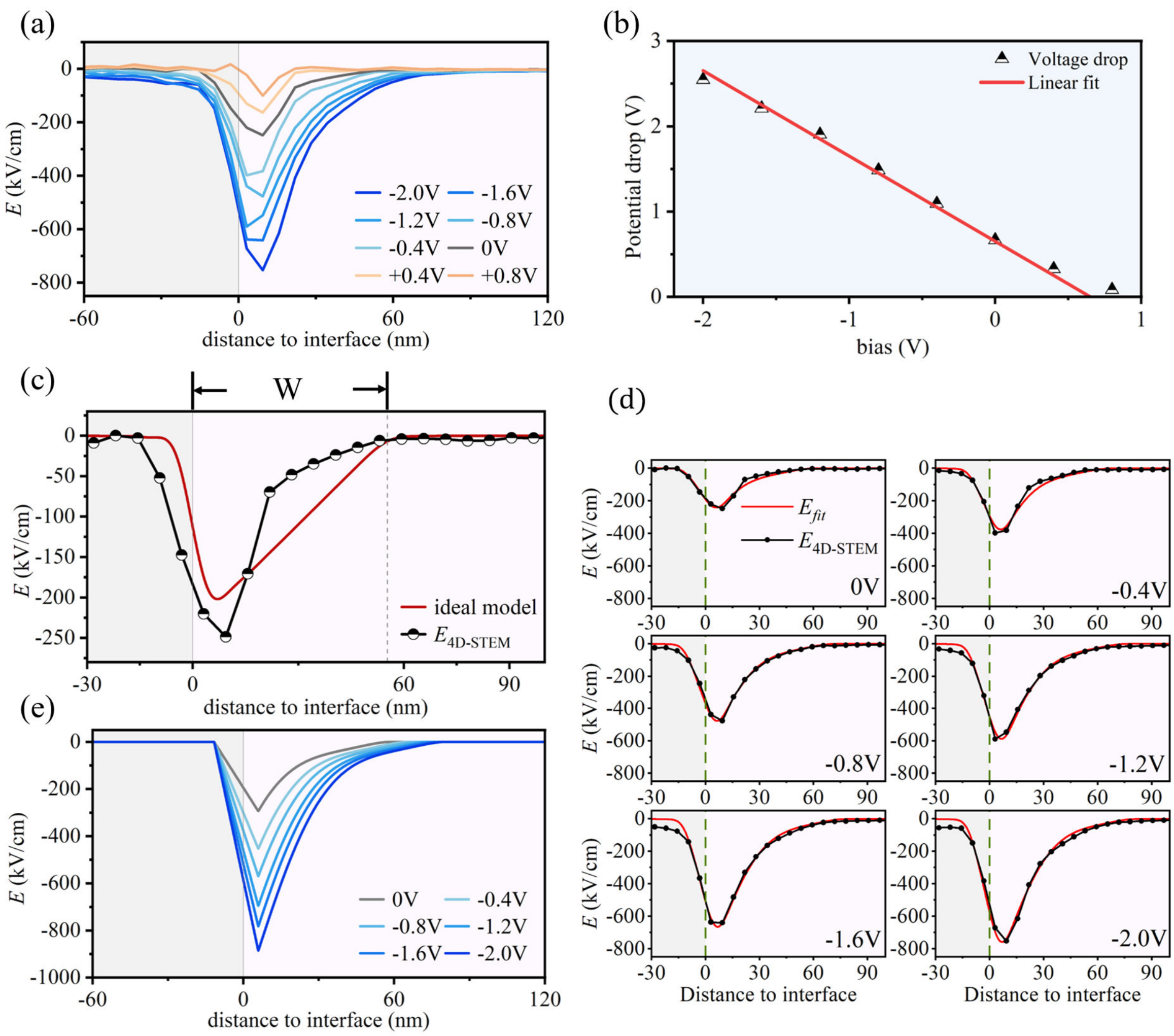

**Figure 6 Quantitative electric field reconstruction.** (a) calculated electric field distributions at different bias from separated $\mathrm{CoM}_x$ using $t_{eff}$; (b) integrated junction potential drop against applied bias; (c) comparison between measured electric field and a convolved ideal Schottky model at zero bias; (d) fitting from forward modelling at different bias; (e) reconstructed original field distribution.

**Table 1 Fitted Schottky parameters from I-V/C-V curves and Cheung-Cheung method.**

| | $n$ | $\varphi_B$ (eV) | $W_0$ (nm) | $R_s$ (MΩ) | $R_p$ (MΩ) |
|---|---|---|---|---|---|
| Bulk device | ~1.1 | ~0.91 | ~63 | - | ~30000 |
| MEMS device | ~2.7 | ~0.67 | ~52 | ~3.4 | ~290 |